\documentclass[prx,preprint,superscriptaddress,amsmath,longbibliography]{revtex4-1}
\usepackage[english]{babel}
\usepackage{graphicx}
\usepackage{color}
\usepackage{float}
\usepackage{graphicx}
\usepackage{amsmath}
\usepackage{amssymb}
\usepackage{gensymb}
\usepackage{multirow}
\usepackage{array}
\DeclareUnicodeCharacter{2009}{\,}

\begin{document}

\title{Space-division multiplexed phase compensation for quantum communication: concept and field demonstration}

\author{Riku Maruyama}
\affiliation{LQUOM Inc., 79-5, Tokiwadai, Hodogaya, Yokohama, 240-8501, Japan}
\affiliation{Yokohama National University, 79-5, Tokiwadai, Hodogaya, Yokohama, 240-8501, Japan}

\author{Daisuke Yoshida}
\affiliation{LQUOM Inc., 79-5, Tokiwadai, Hodogaya, Yokohama, 240-8501, Japan}
\affiliation{Yokohama National University, 79-5, Tokiwadai, Hodogaya, Yokohama, 240-8501, Japan}

\author{Koji Nagano}
\affiliation{LQUOM Inc., 79-5, Tokiwadai, Hodogaya, Yokohama, 240-8501, Japan}

\author{Kouyou Kuramitani}
\affiliation{LQUOM Inc., 79-5, Tokiwadai, Hodogaya, Yokohama, 240-8501, Japan}

\author{Hideyo Tsurusawa}
\affiliation{LQUOM Inc., 79-5, Tokiwadai, Hodogaya, Yokohama, 240-8501, Japan}

\author{Tomoyuki Horikiri}
\affiliation{LQUOM Inc., 79-5, Tokiwadai, Hodogaya, Yokohama, 240-8501, Japan}
\affiliation{Yokohama National University, 79-5, Tokiwadai, Hodogaya, Yokohama, 240-8501, Japan}

\date{\today}

\begin{abstract}
{\bf
Phase-sensitive quantum communication has received considerable attention to overcome the distance limitation of quantum communication.
A fundamental problem in phase-sensitive quantum communication is to compensate for phase drift in an optical fiber channel.
A combination of time-, wavelength-, and space-division multiplexing can improve the phase stability of the optical fiber.
However, the existing phase compensations have used only time- and wavelength-division multiplexing.
Here, we demonstrate space-division multiplexed phase compensation in the Osaka metropolitan networks.
Our compensation scheme uses two neighboring fibers, one for quantum communication and the other for sensing and compensating the phase drift.
Our field investigations confirm the correlation of the phase drift patterns between the two neighboring fibers.
Thanks to the correlation, our space-division multiplexed phase compensation significantly reduces the phase drift and improves the quantum bit error rate.
Our phase compensation is scalable to a large number of fibers and can be implemented with simple instruments.
Our study on space-multiplex phase compensation will support the field deployment of phase-sensitive quantum communication.
}
\end{abstract}

\maketitle

\section*{Introduction}
Quantum communication receives broad attention in quantum cryptography~\cite{bb84,bbm92, Lo2012, Lucamarini2018,Minder2019,Xu2020,Pirandola2020,Liu2021PRL,Pittaluga2021, Chen2021,Zeng2022,Xie2022,Wang2022,Liu2023,Zhu2023,Zhou2023}, distributed quantum computation~\cite{Cirac1999,Yimsiriwattana2004,Jiang2007,Kimble2008,VanMeter2016,Cuomo2020}, and quantum sensor networks~\cite{Jozsa2000,Gottesman2012,Komar2014, Guo2020}.
The extension of quantum communication distance remains a key challenge for realizing a global quantum network~\cite{Kimble2008,Wehner2018,Awschalom2021}.
A single-photon interference scheme is a promising technology to extend the distance limitation in quantum communication~\cite{Duan2001,Lucamarini2018}.
Recent studies have used single-photon interference for twin-field quantum key distribution (QKD)~\cite{Minder2019, Pittaluga2021,Liu2021PRL,Chen2021, Wang2022, Zhou2023, Liu2023}, mode-pairing QKD~\cite{Zeng2022, Xie2022, Zhu2023}, and entanglement distribution between quantum memories~\cite{Delteil2016, Stockill2017, Humphreys2018, Lago-Rivera2021, Pompili2021, Liu2023multinode}.
Since these quantum communication schemes transmit the phase information between distant nodes, the phase fluctuation of the optical fiber channel directly leads to the communication error.
Thus, the phase compensation of the optical fiber plays a crucial role in phase-sensitive quantum communication that uses single-photon interference.

Phase compensation of quantum communication usually monitors the phase drift by a reference laser.
Many studies have exclusively employed a time-division multiplexed phase compensation, which is based on the temporal separation of the signal and reference for implementing phase compensation~\cite{Humphreys2018, Minder2019,Lago-Rivera2021,Liu2021PRL,Chen2021,Pompili2021, Wang2022, Liu2023multinode}.
Recent studies on twin-field QKD have demonstrated significant improvement of the quantum bit error rate (QBER) in a field fiber, where the time-division multiplexed phase compensation is combined with wavelength-multiplexed one~\cite{Pittaluga2021, Clivati2022,Liu2023,Zhou2023}.
While the state-of-the-art phase compensation supports field deployment of phase-sensitive quantum communication, the compensation based on multiplexing the spacial mode still has room for consideration.

We conceptualize a space-division multiplexed phase compensation using two neighboring fibers, one for quantum communication and the other for real-time sensing of the phase drift.
The phase drift in the quantum communication fiber is compensated by the phase drift measured in the reference fiber.
A potential advantage of the space-division multiplexed phase compensation is that a reference fiber can be extended to monitor the phase drifts of multiple nearby fibers for quantum communication.
In that case, one reference fiber can represent the phase drift of the other communication fibers.
This provides scalability of the phase compensation system when the number of fiber usage increases in commercial deployment of phase-sensitive quantum communication.

The validity of the space-division multiplexed phase compensation depends on the correlation of the phase drift patterns in two neighboring fibers.
Ref~\cite{bersin2023} reports the correlating tendency of the phase drifts between two neighboring fibers.
However, it is poorly understood how the phase drift correlation contributes to a space-division multiplexed phase compensation.
Here, we demonstrate space-division multiplexed phase compensation in the Osaka metropolitan network fiber.
Firstly, a space-division multiplexed phase compensation scheme is conceptualized as the differential of the phase drifts between two neighboring fibers.
Then, we investigate the validity of our phase compensation scheme by measuring the phase drift patterns from two neighboring field fibers for over ten days.
Frequency domain analysis confirms the correlation of phase drifts in the two neighboring fibers.
Thanks to the phase drift correlation, our space-division multiplexed phase compensation significantly improves the QBER in the data post-processing.
Furthermore, we investigate the impact of the fiber environment (buried or aerial) on the QBER.
Our space-division multiplexed phase compensation improves the QBER of both buried and aerial fibers, although the buried fibers have higher stability of the compensated QBER for several days.
Our conceptualization and demonstration of space-division multiplexed phase compensation will broaden the opportunities for the field deployment of phase-sensitive quantum communication.

\section*{Results}
\subsection*{Experimental setup}
\begin{figure}[h]
    \centering
    \includegraphics[width=0.5\textwidth]{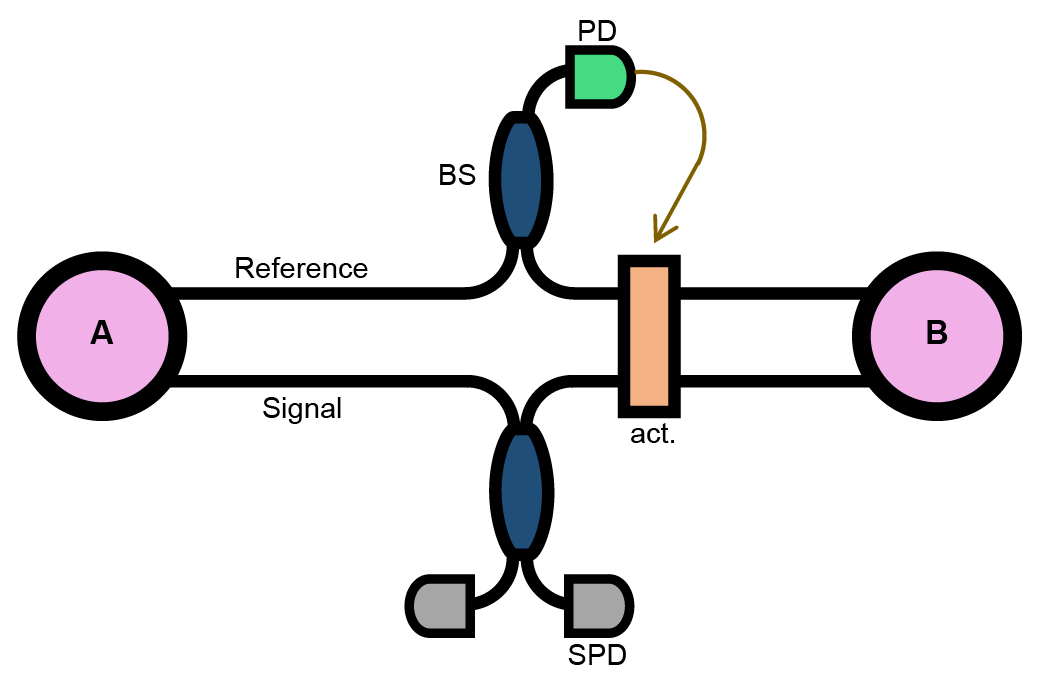}
    \caption{
    {\bf Our scheme of space-division multiplexed phase compensation.}
    Our compensation scheme prepares two neighboring fibers, one for the quantum communication (Signal) and the other for the phase compensation (Reference).
    Nodes A and B represent communication nodes.
    The Reference fiber monitors the phase drift to compensate for the phase drift in the Signal fiber either with real-time feedback (brown arrow) or in data post-processing. 
    The actuator is used for real-time feedback.
    BS, beam splitter; PD, a photodetector for phase monitoring; act., common actuator for both paths; SPD, single photon detector for quantum communication.
    }
    \label{fig:scheme}
\end{figure}

We conceptualize our phase compensation scheme in Fig.~\ref{fig:scheme}.
The space-division multiplexed phase compensation prepares two neighboring fibers, one fiber for quantum communication (signal fiber) and the other for real-time phase-sensing (reference fiber).
The phase drift in the reference fiber provides information on the phase drift of the signal fiber, assuming the correlation of the phase drifts in the two fibers (Fig.~\ref{fig:scheme}).
We can implement the space-division multiplexed phase compensation scheme either with real-time feedback or in data post-processing.
One can use the common actuator shown in Fig.~\ref{fig:scheme} for real-time feedback implementation.
Phase compensation in data post-processing is often valid for single-photon interference schemes~\cite{Liu2021PRL, Chen2021, Wang2022, Liu2023, Zhou2023}.
A fundamental question of the space-division multiplexed compensation is the correlation of phase drift patterns between the signal and reference fibers in a field environment, especially in a metropolitan area.
To address this point, we investigate the phase drift pattern of commercial fibers in the Osaka metropolitan area.

\begin{figure}[h]
    \centering
    \includegraphics[width=0.95\textwidth]{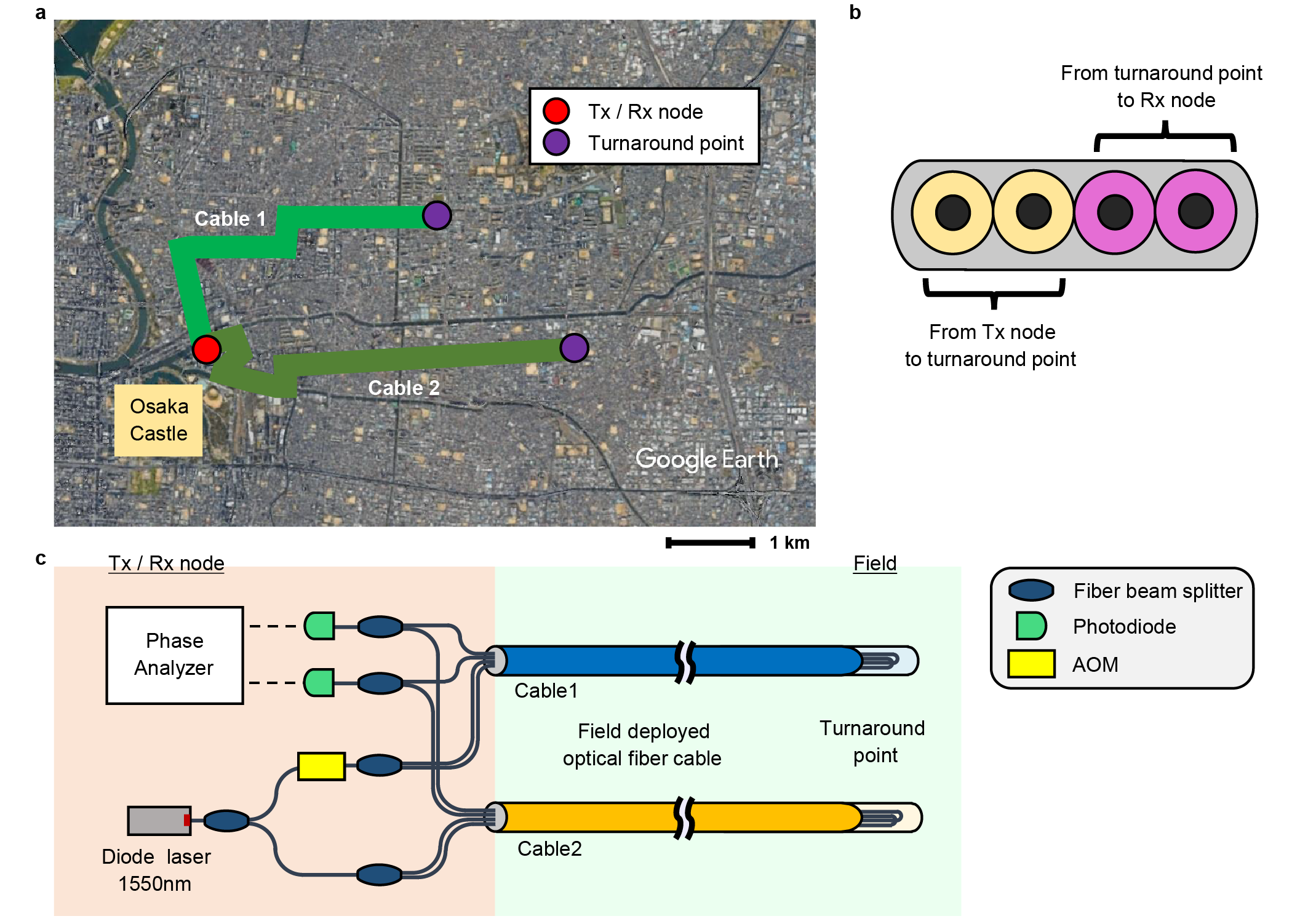}
    \caption{
    {\bf Experimental setup in the Osaka metropolitan fiber network.}
    {\bf a}, Map of our first setup in the Osaka metropolitan area.
    Each of the two buried fibers starts from a central building to a turnaround point.
    See Table~$1$ for detailed information on the fibers.
    {\bf b}, We used four-fiber ribbon cables.
    Reference fiber is the nearest neighbor of Signal fiber.
    {\bf c}, Our optical setup. Whole fibers are single-mode fibers. AOM represents an acousto-optic modulator.
    }
    \label{fig:setup}
\end{figure}

Figure~\ref{fig:setup} explains our experimental setup using data post-processing.
We installed our measurement system at a building near Osaka Castle, a central part of the Osaka metropolitan area (Fig.~\ref{fig:setup}a).
Cable~$1$ goes through a central area of Osaka and returns at the turnaround point to the starting point (Figs.~\ref{fig:setup}a and b).
Similarly, Cable~$2$ starts from the same building and returns at another turnaround point (Figs.~\ref{fig:setup}a and b).
Cable~$1$ and Cable~$2$ are entirely buried fibers (see Table~\ref{table:1}).
We used four-fiber ribbon in both Cable~$1$ and $2$.
The signal fiber is the nearest neighbor of the reference fiber (Fig.~$2$c).
Hence, we have prepared the commercial fibers to investigate the correlation of the phase drift between the two neighboring fibers.
A heterodyne interferometric measurement detects the phase drifts of the signal and reference fibers 
in a sub-wavelength precision (Fig.~\ref{fig:setup}a and Methods).
We obtain a differential phase by subtracting the phase drift of the reference fiber from that of the signal fiber in post-processing.
We emphasize that our measurement of the phase drifts continues for $7$~days.
Our experimental time window is sufficiently long to evaluate the long-term robustness of the space-division multiplexed phase compensation in a metropolitan fiber network.

\subsection*{Phase compensation of buried fibers}
\begin{figure}[h]
    \centering
    \includegraphics[width=0.95\textwidth]{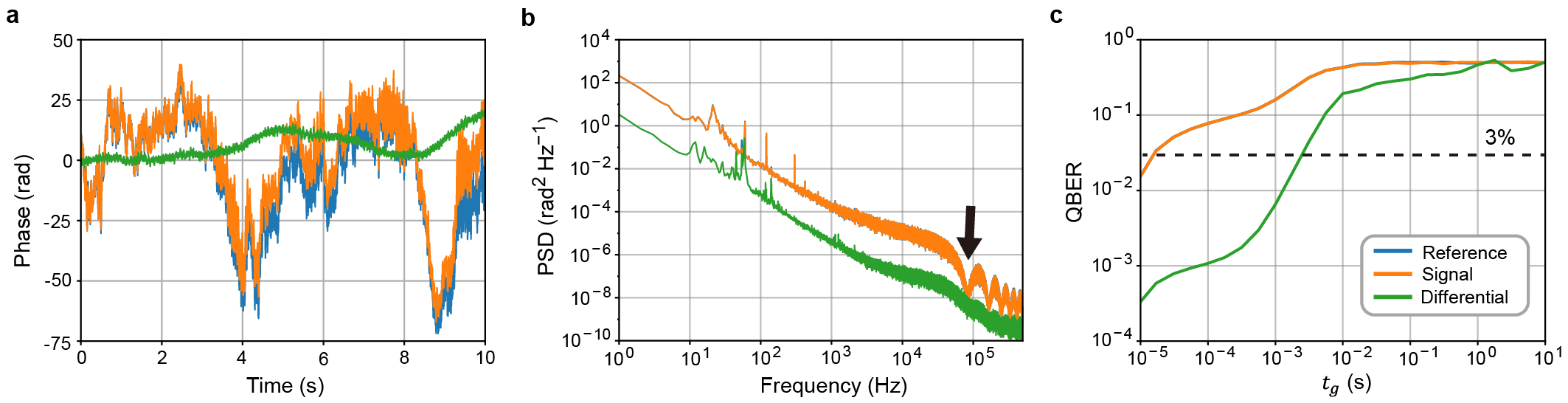}
    \caption{
    {\bf Phase drift compensation of buried fibers.}
    {\bf a}, A typical dataset of the phase drift in the buried fibers with an interval of $10$~s.
    The differential in the phase drift between the Reference and Signal fiber is computed.
    {\bf b}, PSD is computed from the phase dataset of {\bf a}.
    The arrow indicates the first dip of the PSD of the Reference fiber.
    {\bf c}, QBER is computed from the phase dataset of {\bf a} (see also Methods and Supplemental Fig.~\ref{supl:qber_analysis}).
    In {\bf b} and {\bf c}, the results of the Reference and Signal are almost identical and overlap in the plots.
    For the definition of $t_g$, refer to the Methods section.
    }
    \label{fig:buried_static}
\end{figure}

Figure~\ref{fig:buried_static} analyzes the phase drift measured in the buried fibers.
The phase drifts of the reference and signal fibers are overlayed in Fig.~\ref{fig:buried_static}a.
Fig.~\ref{fig:buried_static}b compares the power spectral density (PSD) of the phase as a function of frequency ranging from $1$~Hz to $5 \times 10^5$~Hz.
The PSD of the reference fiber is correlated with that of the signal fiber.
Due to the phase drift correlation, the baseline of the PSD of the differential phase is approximately $100$~times lower than that of the signal fiber.
These results validate our space-division multiplexed phase compensation between the two neighboring fibers.

We evaluate the impact of our space-division multiplexed phase compensation on quantum communication.
Fig.~\ref{fig:buried_static}c shows the QBER computed from the phase measurements.
As expected, our phase compensation improves the QBER of the signal fiber.
Here, we evaluate the interval until the QBER reaches $3$\% in the absence of phase compensation, $t_{0.03}$.
Without and with the space-division multiplexed phase compensation, $t_{0.03}$ is $1.4\times10^{-5}$~s and $2.5\times10^{-3}$~s, respectively.
Thus, our space-division multiplexed phase compensation improves the interval $t_{0.03}$ by a factor of $180$.

Moreover, we investigate the long-term robustness of our space-division multiplexed phase compensation performance in the buried fibers.
(See Supplemental Figs.~\ref{supl:buried_psd} and \ref{supl:buried_qber} for the full results of the phase data in spectrograms and estimated QBERs for $7$~days, respectively.)
Figure~\ref{fig:buried_dynamic} summarizes the time variation of the interval $t_{0.03}$.
Both the measured and compensated data are almost constant.
Our space-division multiplexed phase compensation robustly improves the interval $t_{0.03}$ for $7$~days.
Considering the simple setup of our compensation scheme, this robustness and improvement should be notable.

\subsection*{Comparison between buried and aerial fibers}
A field fiber network usually combines buried fibers and aerial fibers.
Thus, our space-division multiplexed phase compensation should be investigated using fibers, including aerial parts.
We have prepared another fiber setup in the Osaka metropolitan fiber network as shown in Supplemental Fig.~\ref{supl:aerial_map}.
The field fibers in the second setup consist mostly of aerial and partially buried fibers (for detail, see the caption of Table~\ref{table:1}).
Hereafter, these fibers are simply called aerial fibers unless it is confusing.
We have measured the phase of the two neighboring fibers for $5$~days.

\begin{figure}[h]
    \centering
    \includegraphics[width=0.95\textwidth]{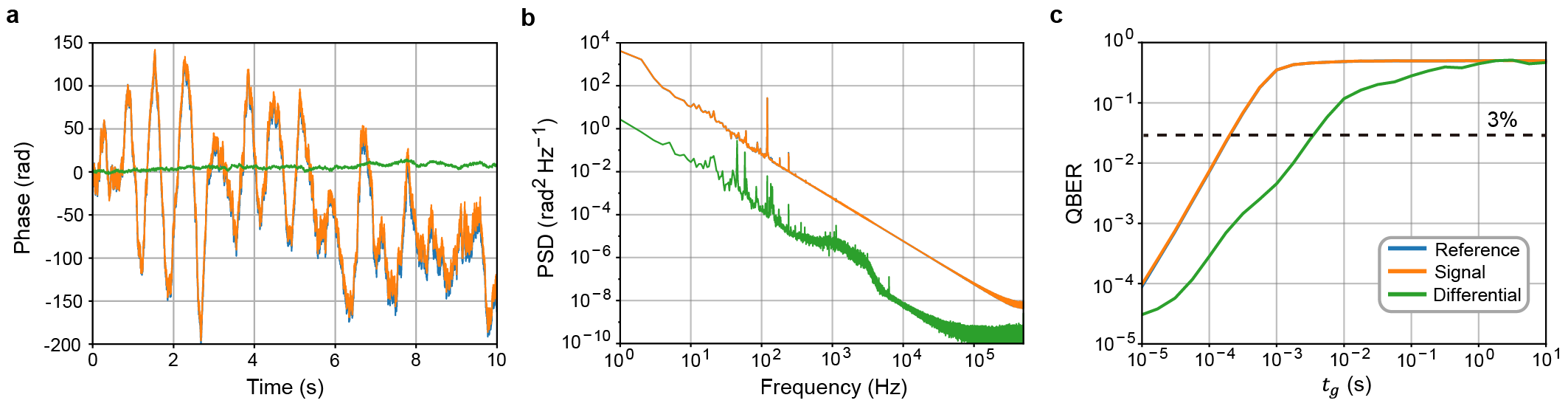}
    \caption{
    {\bf Phase drift compensation of aerial fibers.}
    {\bf a}, A typical data of the phase drift in the aerial fibers with an interval of $10$~s.
    The map of the fibers in our second setup is available in Supplemental Fig.~\ref{supl:aerial_map}.
    PSD ({\bf b}) and QBER analysis ({\bf c}) are computed from the raw data of {\bf a}.
    The results of the Reference and Signal are almost identical and overlap in the plots.
    For the definition of $t_g$, refer to the Methods section.
    }
    \label{fig:aerial_static}
\end{figure}

Figure~\ref{fig:aerial_static}a shows the phase drift measured within an interval of $10$~s.
Although the phase of the aerial fiber obviously fluctuates at a few Hz, the phase drift pattern shows an apparent smilarity between the reference and signal fibers.
In Fig.~\ref{fig:aerial_static}b, the PSD analysis confirms the correlation of the phase drift between the two fibers.
Indeed, the space-division multiplexed phase compensation decreases the baseline of the phase PSD by a factor of approximately $100$ (see curves labeled as Reference and Differential in Fig.~\ref{fig:aerial_static}b).
Figure~\ref{fig:aerial_static}c shows the QBER of the reference and compensated phase.
Again, we set the QBER threshold at $3$\% and compute the interval $t_{0.03}$ of the aerial fiber.
Our phase compensation improves the interval $t_{0.03}$ from $2.5\times 10^{-4}$~s (without compensation) to $4.4\times 10^{-3}$~s (with compensation) in data post-processing.
These results validate our space-division multiplexed phase compensation, even in aerial fibers.

We also check the long-term robustness of our space-division multiplexed phase compensation in aerial fibers.
(See Supplemental Figs.~\ref{supl:aerial_psd} and \ref{supl:aerial_qber} for the full results of the spectrograms and estimated QBERs of the aerial fibers for $5$~days.)
In the aerial fibers, the phase drift fluctuates in a daily cycle.
The interval of $t_{0.03}$ without compensation ranges from $2.5\times 10^{-5}$~s to $4.4\times 10^{-4}$~s, whose fluctuation amplitude is significantly larger than that of buried fibers shown in Fig.~\ref{fig:buried_dynamic}.
Nonetheless, our space-division multiplexed phase compensation scheme has improved the interval $t_{0.03}$ of the aerial fibers by a factor of $18$ in our experiment.

Moreover, we find that the field environment (buried or aerial) affects the low-frequency phase drift.
The aerial and buried fibers have unique peaks at $\sim 2$~Hz (Figs.~\ref{fig:aerial_static}a and b) and $\sim 20$~Hz (Fig.~\ref{fig:buried_static}b), respectively.
These peaks indicate (i) wind galloping on the aerial fiber~\cite{Wuttke2003, Ding2017} and (ii) mechanical vibration from transportation to the buried fiber~\cite{HAO2001}.
Although the low-frequency phase drifts have little impact on our results of $t_{0.03} \sim 10^{-3}$~s, understanding the phase drift in a full frequency range may be helpful for the further improvement of QBER in a field fiber network.

\subsection*{Technical limitations of our setups}
Our field study has demonstrated that space-division multiplexed phase compensation is applicable in both buried and aerial fiber conditions.
Now, we address the technical limitations of our setups for the future improvement of space-division multiplexed phase compensation.
In the first setup (buried fibers), the compensation has improved the interval $t_{0.03}$ to $2.5\times10^{-3}$~s (Fig.~\ref{fig:buried_static}c).
This means that the high-frequency noise above $\sim 10^3$~Hz limits the further improvement of QBER.
In the PSD, the reference and signal fiber phase noises periodically dip at $83$ kHz and its higher harmonics (Fig.~\ref{fig:buried_static}b).
These dips are explained as the laser frequency noise cancellation because their frequency matches the theoretical prediction (Methods and Table~\ref{table:1}).
It will be the case when the lasers used for the signal and reference fibers are the same (like our demonstration setup) or as long as they are synchronized.
Therefore, the reduction of the laser frequency noise for the reference measurement should improve QBER.

In the second setup (aerial fibers), the compensated QBER shows a significant fluctuation in a daily cycle.
We question if the ``precision'' of our phase compensation also fluctuates in the aerial fiber condition.
Supplemental Fig.~\ref{supl:visibility} shows the long-term change in the interference visibility of our system.
The visibility of the second setup significantly fluctuates in comparison with that of the buried fiber.
Thus, the visibility fluctuation in the second setup comes from the aerial part of the fiber.
The fluctuation of polarization in the aerial fibers is a sensible explanation for the visibility fluctuation in the second setup.
This explanation is supported by previous field investigations on the polarization drift of a field fiber \cite{bersin2023}, where the polarization drift couples with weather conditions (e.g., wind speed and temperature).
Hinted from the previous reports, we compare the visibility fluctuation with the weather data of the Osaka metropolitan area.
The visibility of the second setup tends to become unstable when the wind speed is relatively high.
This tendency indicates that the weather conditions might change the interference visibility of the second setup.
Taken together, the additional compensation for polarization drift may improve the long-term robustness of our space-division multiplexed phase compensation, especially when the fiber is in an aerial environment.

\section*{Discussion}
Phase-sensitive quantum communication has been developed in a laboratory~\cite{Chou2005,Delteil2016,Stockill2017, Humphreys2018,Lago-Rivera2021,Pompili2021,Pittaluga2021,Wang2022,Liu2023, Zhu2023}, which provides a relatively stable environment for the optical fiber.
Phase compensation of field fibers is a fundamental problem as the proven technologies are going toward field deployment.
While previous field studies have mainly employed time- and wavelength-division multiplexed phase compensation~\cite{Clivati2022,Liu2021PRL,Chen2021, Liu2023multinode}, space-division multiplexed compensation still has room for consideration.
Our study has demonstrated space-division multiplexed phase compensation of both aerial and buried fibers in a metropolitan network for a week scale.
The space-division multiplexed phase compensation improves $t_{0.03}$ up to $2.5$ ms in the Osaka metropolitan area fiber network.
We emphasize that our space-division multiplexed phase compensation is complementary to the time- and wavelength-division ones.
Thus, the combination of time-, wavelength-, and space-division multiplexed phase compensation should broaden the opportunities for the field deployment of phase-sensitive quantum communication.

\begin{figure}[h]
    \centering
    \includegraphics[width=0.95\textwidth]{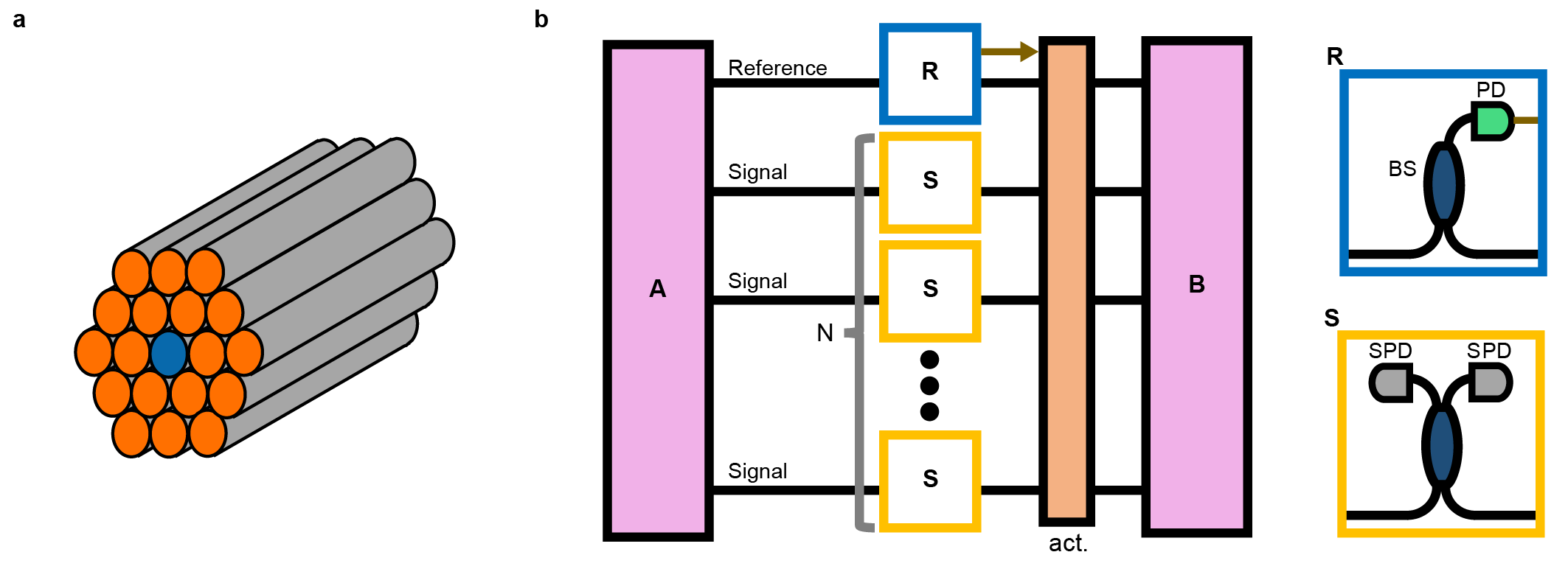}
    \caption{
    {\bf Scalability of phase compensation system.}
    We consider the application of the space-division multiplexed phase compensation scheme to the fiber bundle.
    Assuming the phase drift correlation between the $N+1$ fibers, one reference system can provide real-time feedback (brown arrow) on the phase drift to the other communication fibers at once.
    This leads to the scalability of the phase compensation system when using the fiber bundle for large-scale quantum communication.
    }
    \label{fig:efficiency}
\end{figure}

Finally, we discuss the system scalability of our space-division multiplexed phase compensation.
Phase-sensitive quantum communication receives intensive expectations towards field deployment and commercial roll-out~\cite{Chen2021,Clivati2022,bersin2023, Liu2023multinode}.
In that case, many fibers will be used in parallel so that the total capacity of quantum communication increases.
Time- and wavelength-division multiplexed phase compensation suppose that a compensation system individually senses and compensates the phase drift for every fiber.
Thus, the optical system of the fiber-by-fiber phase compensation schemes may become gigantic as the number of fiber usage increases.
In contrast, our space-division multiplexed phase compensation may be able to sense the phase drift of multiple fibers by using only one reference fiber (Fig.~\ref{fig:efficiency}).
In other words, only one optical system can compensate for the phase drifts of many fibers at once.
Therefore, our space-division phase compensation will offer the system scalability as a unique advantage for the roll-out of phase-sensitive quantum communication.

\clearpage
\section*{References}
\bibliography{ref}
\bibliographystyle{naturemag2.bst}

\clearpage
\noindent
\section*{Methods}
\subsection*{Experimental setup of the phase measurement}
Figure~\ref{fig:setup}(c) shows the schematic of the phase measurement.
We use an external cavity diode laser (RIO ORION) with a wavelength of $1550\pm2$~nm and a Lorentzian linewidth of $<1$~kHz.
A beam splitter just after the laser source splits the laser beam for two cables.
An acousto-optic modulator device shifts the frequency of the laser light incident to one cable by $40$~MHz for the heterodyne interferometry.
The laser light is split to two paths (for signal and reference) before the cable input.
A data logger with a sampling rate of $1$~MSa/s collected a phase dataset for $10$~s and paused the data collection for $50$~s as shown in Supplemental Fig.~\ref{supl:qber_analysis}.
We repeated the data collection cycles in the first setup for $7$~days and in the second setup for $5$~days.
The total data of $1.4$~TB is analyzed.

Table~\ref{table:1} summarizes the one-way lengths of buried and aerial fibers of each of the cables.
$\Delta{L}$ is calculated as $|L_1 - L_2| \times 2$, where $L_1$ and $L_2$ are the one-way length of a cable and its complementary cable in a setup, respectively.
To explain the dip of the high-frequency noise in Fig.~\ref{fig:buried_static}b, the frequency interval ($\Delta{f}$) is calculated as $\Delta{f} = c / n\Delta{L}$, where $c$ and $n$ are the speed of light in vacuum and the refractive index of the optical fiber.
Note that we set the coefficients as $c = 3.0 \times 10^8$~m/s and $n = 1.5$.
In our phase measurement, laser frequency noise is canceled at every $\Delta{f}$ along the frequency axis.
In the first setup, the calculation of $\Delta{f}$ shows quantitative agreement with the experimental result of Fig.~\ref{fig:buried_static}b.
In the second setup, the frequency interval of $\Delta{f}$ is out of our measurement range (Fig.~\ref{fig:aerial_static}b).

\subsection*{Calculation of quantum bit error rate (QBER)}
Here, we consider the single photon interference for the quantum communication with an interferometer having two output ports.
When the interference condition is set to be constructive for one output port, the photon is always detected in that port.
The probability of the photon detection in the other port increases when the interference condition is shifted.
This interference condition shift leads to the communication error.
QBER is an error metric from the point of view of QKD~\cite{Clivati2022}.
Here, we ignore the other factors increasing QBER, such as the non-unity quantum efficiency of the detector.
QBER in a given measurement time window is formulated as
\begin{equation}
{\rm QBER} = \frac{1}{n} \sum_{i = 1}^{n} \sin^2{\frac{\phi_i}{2}},
\label{eq:discrete}
\end{equation}
where $\phi_i$~$(i = 1, 2, 3,..., n)$ is the round-trip phase difference of the two arms of the interference, and $n$ is the number of phase data points.
Here, $\phi_i = 0$ is defined as the constructive phase condition explained above.

\begin{figure}[h]
    \centering
    \includegraphics[width=0.5\textwidth]{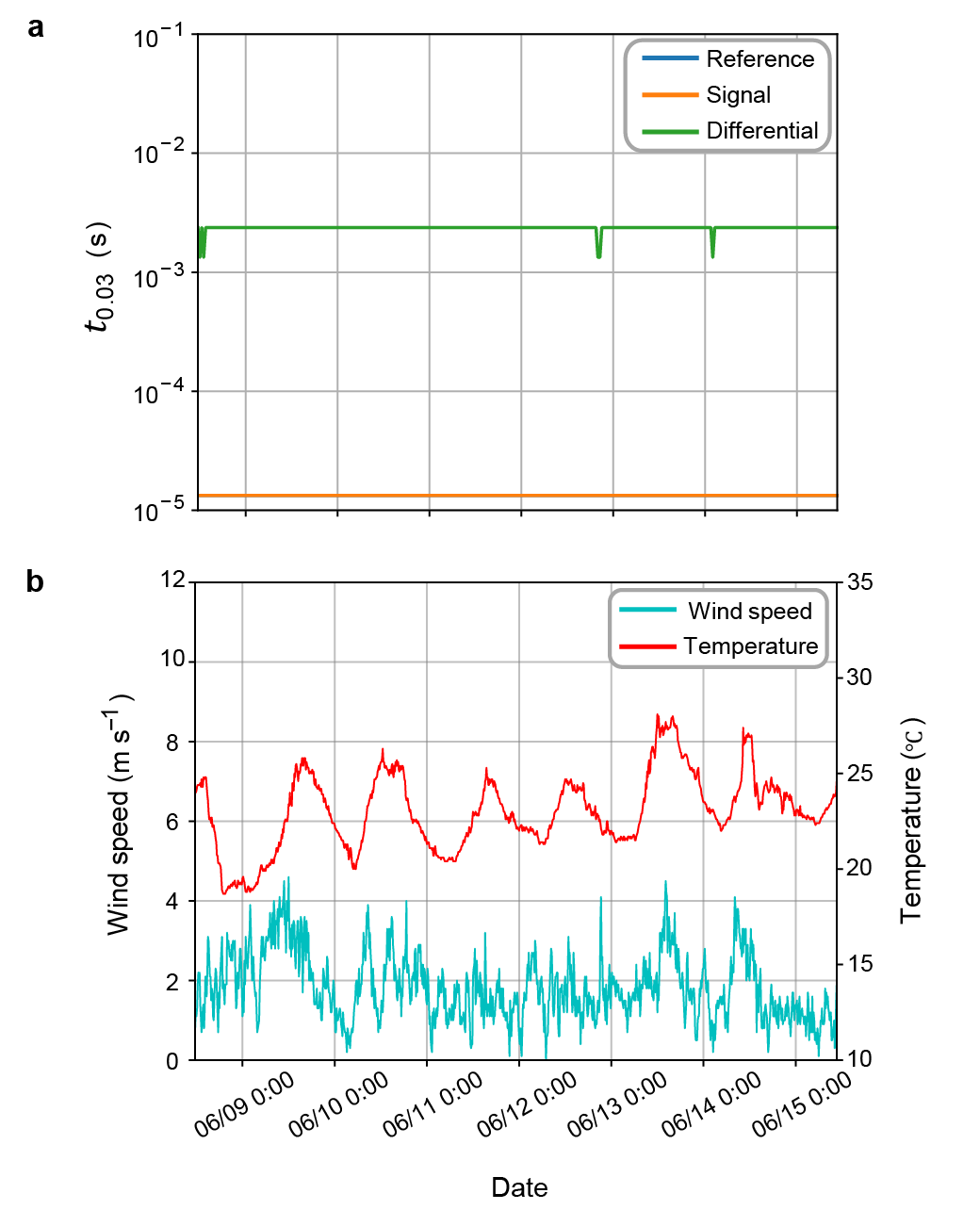}
    \caption{
    {\bf Long-term data of QBER in buried fibers.}
    {\bf a}, The characteristic interval of $t_{0.03}$ is computed as a function of time from the series of QBER data with a threshold of QBER of $3$\%.
    Our experimental time window is $7$~days.
    The full dataset of QBER is available in Supplemental Fig.~\ref{supl:buried_qber}.
    The result of the Reference and Signal are almost identical and overlap in the plot.
    {\bf b}, Weather data of the Osaka metropolitan area. The weather data source is Japan Meteorological Agency~\cite{JMA}.
    }
    \label{fig:buried_dynamic}
\end{figure}

\begin{figure}[h]
    \centering
    \includegraphics[width=0.5\textwidth]{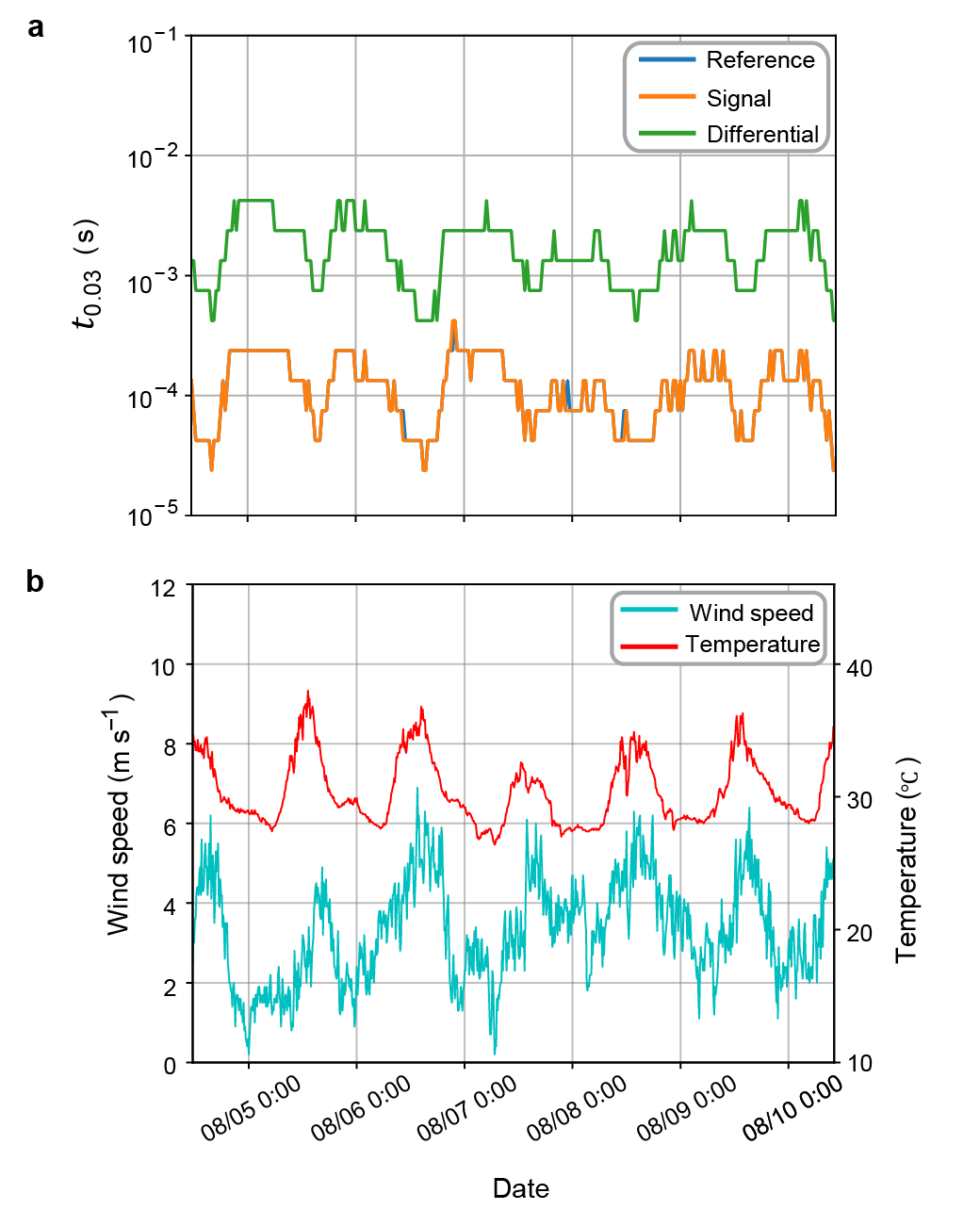}
    \caption{
    {\bf Long-term fluctuation of QBER in aerial fibers.}
    {\bf a}, The time interval of $t_{0.03}$ for QBER is computed as a function of time.
    The full dataset is available in Supplemental Fig.~\ref{supl:aerial_qber}.
    Our experimental time window is $5$~days.
    The result of the Reference and Signal are almost identical and overlap in the plot.
    {\bf b}, Weather data of the Osaka metropolitan area. The weather data source is Japan Meteorological Agency~\cite{JMA}.
    }
    \label{fig:aerial_dynamic}
\end{figure}

In our study, one phase dataset has a duration of $10$~s with a sampling rate of $1$~MSa/s.
The dataset is split into subsets so that one subset corresponds to a duration of $t_{\rm g}$ (see Supplemental Fig.~\ref{supl:qber_analysis}).
We follow the equation~(\ref{eq:discrete}) to compute QBER from a phase data subset, where $t_{\rm g} = 10^{-5 + k/4}$~s and $k = 0, 1, 2, ..., 24$.
Then, QBER is averaged over the subsets in a phase dataset.
In Figs.~\ref{fig:buried_dynamic} and \ref{fig:aerial_dynamic}, QBER is additionally averaged over phase datasets (Supplemental Fig.~\ref{supl:qber_analysis}).
In our analysis, $t_{0.03}$ corresponds to the time when ${\rm QBER} (t_{\rm g})$ exceeds $3$\%.

\clearpage
\noindent
{\bf Acknowledgements}\\
The authors thank Yoshitaro Tsuji, Motoki Tanaka, Soichiro Nishiuma, and Takaaki Yui (OPTAGE Inc.) for arranging the field fiber experiment and Mahoro Yoshida for supporting our project.
This research was supported by Deep-Tech Startups Support Program (New Energy and Industrial Technology Development Organization, Japan).
\\

\noindent
{\bf Author contributions}\\
D.Y. and K.N. conceptualized the study.
R.M., D.Y., K.N., and K.K. performed experiments.
R.M. and K.N. analyzed the data.
R.M., D.Y., K.N., and H.T. wrote the manuscript.
All authors discussed the results and commented on the manuscript.
T.H. supervised the project.
\\

\noindent
{\bf Competing interests}\\
All authors have a financial interest in LQUOM Inc.
\\

\noindent
{\bf Materials \& Correspondence}\\
Correspondence and material requests should be addressed to D.Y.

\clearpage

\clearpage

\clearpage

\clearpage

\clearpage

\clearpage

\clearpage

\clearpage
\setcounter{table}{0}
\renewcommand{\tablename}{\textbf{Table}}
\renewcommand{\thetable}{\arabic{table}}
\begin{table}[]
\caption{
{\bf Parameters of the field fibers.
}
We prepared Cable~$1$ and $2$ in the first setup (Fig.~\ref{fig:setup}a) and Cable~$3$ and $4$ are prepared in the second setup (Supplemental Fig.~\ref{supl:aerial_map}). 
Cables 1 and 2 are composed entirely of buried fiber, while Cables 3 and 4 consist of approximately 70\% aerial fiber, with the remainder being buried fiber.
}
\begin{tabular}{ccccc}
\hline
 & \begin{tabular}[c]{@{}l@{}}Round trip\\length (km)\end{tabular} & $\Delta{L}$ (km) & $\Delta{f}$ (Hz)  \\
\hline
Cable $1$   &$10.4$ &\multirow{2}{*}{$2.4$}   & \multirow{2}{*}{$8.3\times 10^4$} \\
Cable $2$   & $12.8$ &   &   \\
\hline
Cable $3$   & $11.5$ & \multirow{2}{*}{$0.0$} & \multirow{2}{*}{-} \\
Cable $4$   & $11.5$ &   & \\
\hline
\end{tabular}
\label{table:1}
\end{table}

\setcounter{figure}{0}
\renewcommand{\figurename}{\textbf{Supplemental Fig.}}
\renewcommand{\thefigure}{\arabic{figure}}

\clearpage
\begin{figure}[h]
    \centering
    \includegraphics[width=0.95\textwidth]{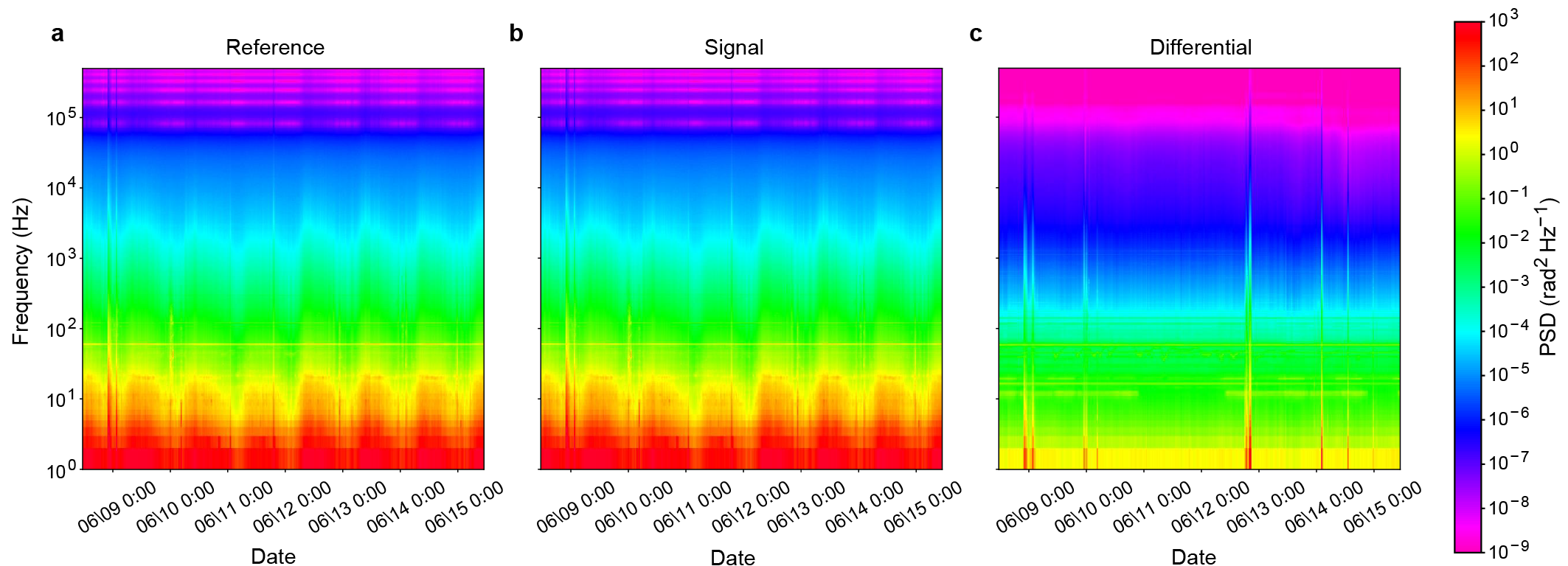}
    \caption{
    {\bf Long-term spectrogram of PSD data collected in buried fibers.}
    Spectrogram of the reference fiber {\bf a}, the signal fiber {\bf b}, and the differential {\bf c} are computed from the full-time data.
    One horizontal line shows a PSD with an interval of $30$~minutes.
    A sudden drop of QBER at 06/09 AM 01:43 is caused by an event of a weak earthquake.
    }
    \label{supl:buried_psd}
\end{figure}

\clearpage
\begin{figure}[h]
    \centering
    \includegraphics[width=0.95\textwidth]{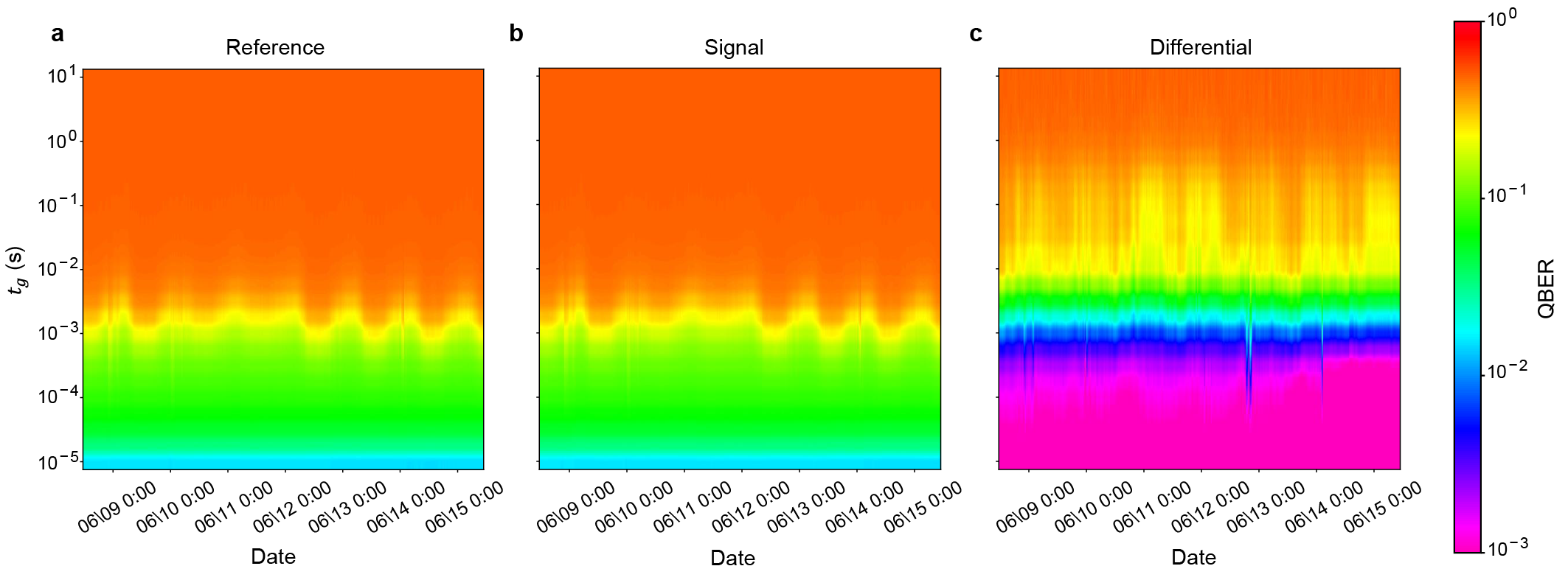}
    \caption{
    {\bf Long-term spectrogram of QBER collected in buried fibers.}
    Spectrogram of QBER is computed from the full-time data collected in the reference fiber {\bf a}, the signal fiber {\bf b}, and the differential {\bf c}.
    }
    \label{supl:buried_qber}
\end{figure}

\clearpage
\begin{figure}[h]
    \centering
    \includegraphics[width=0.95\textwidth]{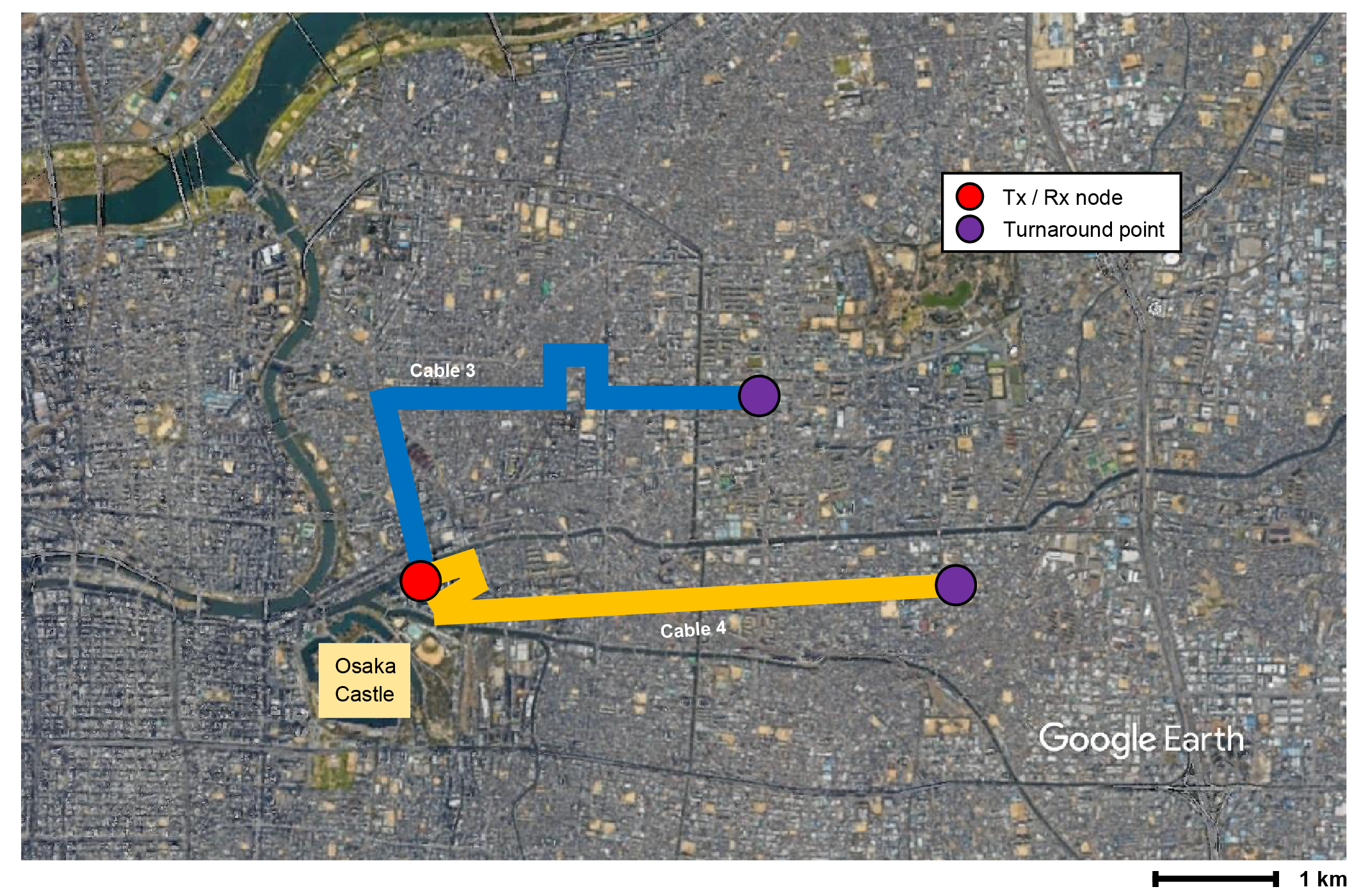}
    \caption{
    {\bf Map of the second setup.}
    Both two cables combine aerial and buried fibers.
    The detailed information is available in Table~$1$.
    The routes of the second setup are similar but partly different from those of the first setup.
    }
    \label{supl:aerial_map}
\end{figure}

\clearpage
\begin{figure}[h]
    \centering
    \includegraphics[width=0.95\textwidth]{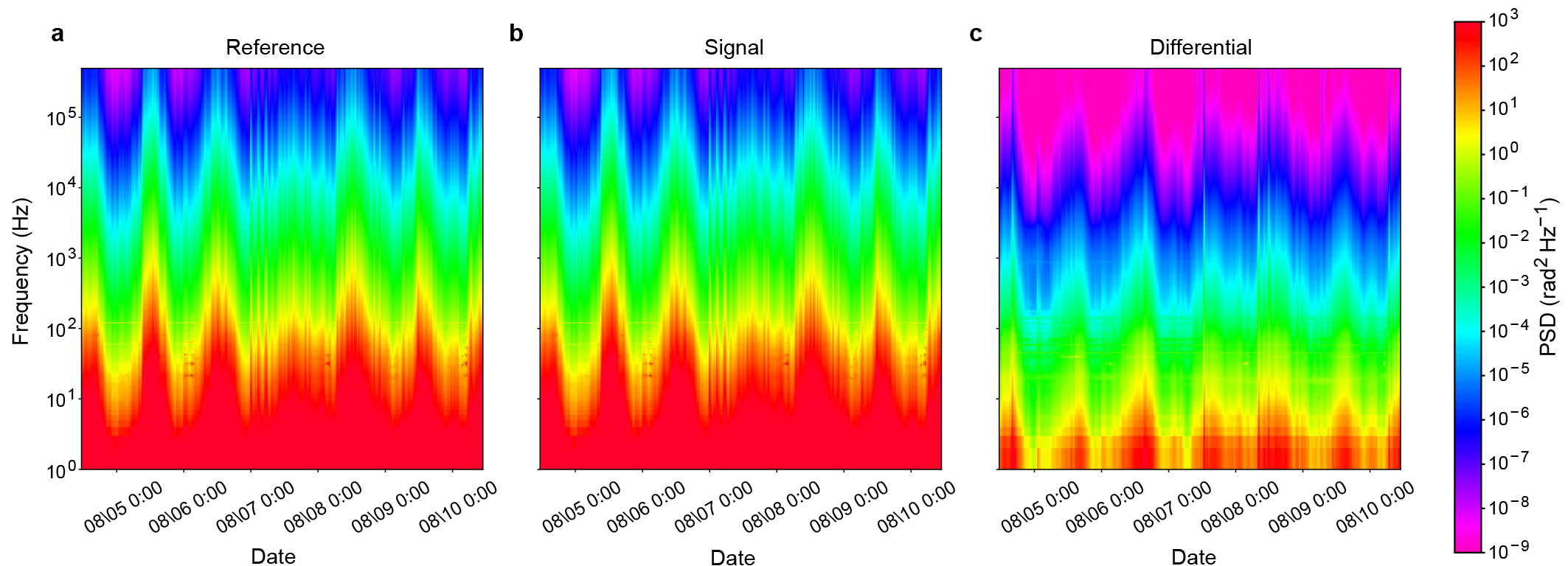}
    \caption{
    {\bf Long-term spectrogram of PSD data collected in aerial fibers.}
    Spectrogram of PSD of the reference fiber {\bf a}, the signal fiber {\bf b}, and the differential {\bf c} are computed from the full-time data.
    }
    \label{supl:aerial_psd}
\end{figure}

\clearpage
\begin{figure}[h]
    \centering
    \includegraphics[width=0.95\textwidth]{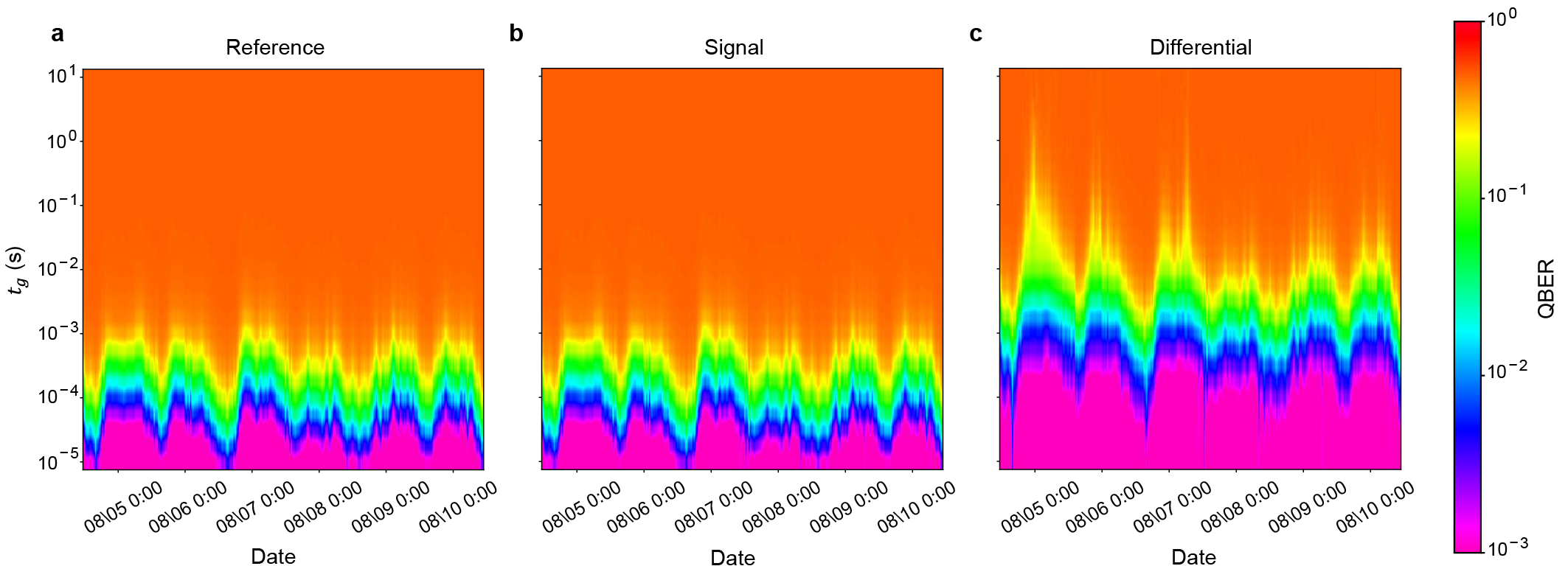}
    \caption{
    {\bf Long-term spectrogram of QBER data collected in aerial fibers.}
    Spectrogram of QBER is computed from the full-time data collected in the reference fiber {\bf a}, the signal fiber {\bf b}, and the differential {\bf c}.
    }
    \label{supl:aerial_qber}
\end{figure}

\clearpage
\begin{figure}[h]
    \centering
    \includegraphics[width=0.95\textwidth]{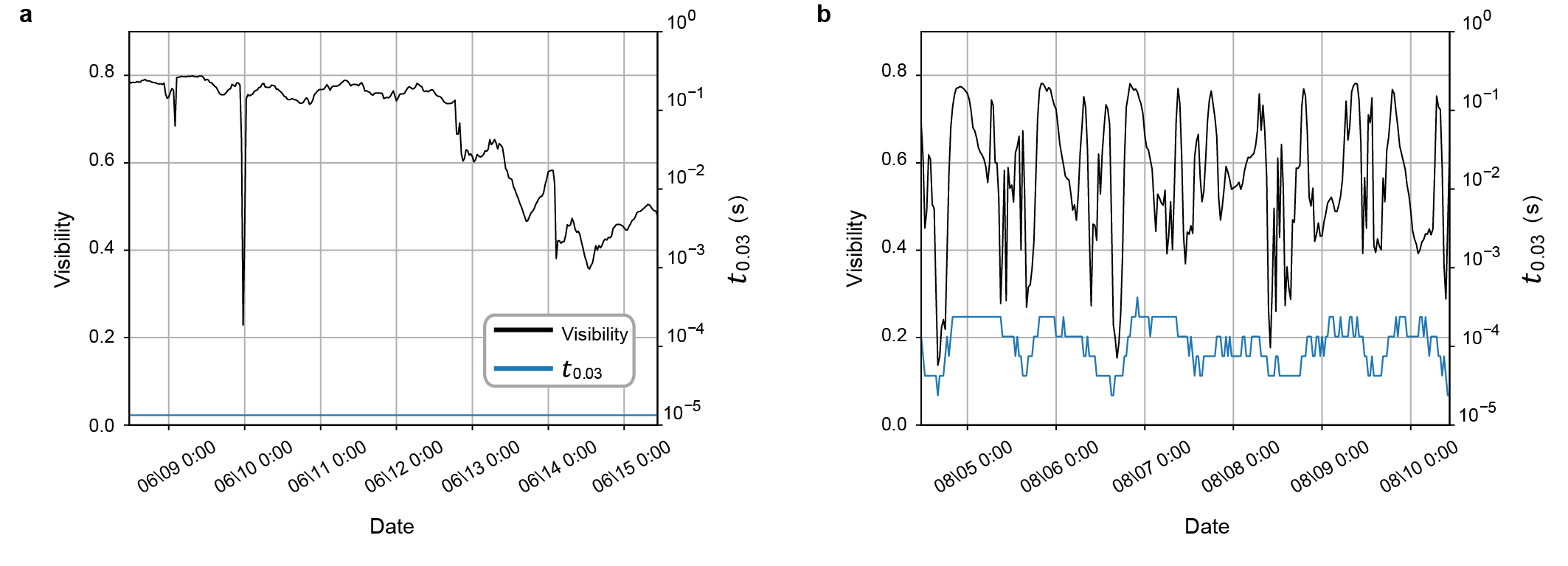}
    \caption{
    {\bf Long-term fluctuation of the interference visibility in our setups.}
    {\bf a}, buried fibers. The visibility of the interferometer and $t_{0.03}$ for the reference fiber are shown.
    {\bf b}, aerial fibers.
    }
    \label{supl:visibility}
\end{figure}

\clearpage
\begin{figure}[h]
    \centering
    \includegraphics[width=0.95\textwidth]{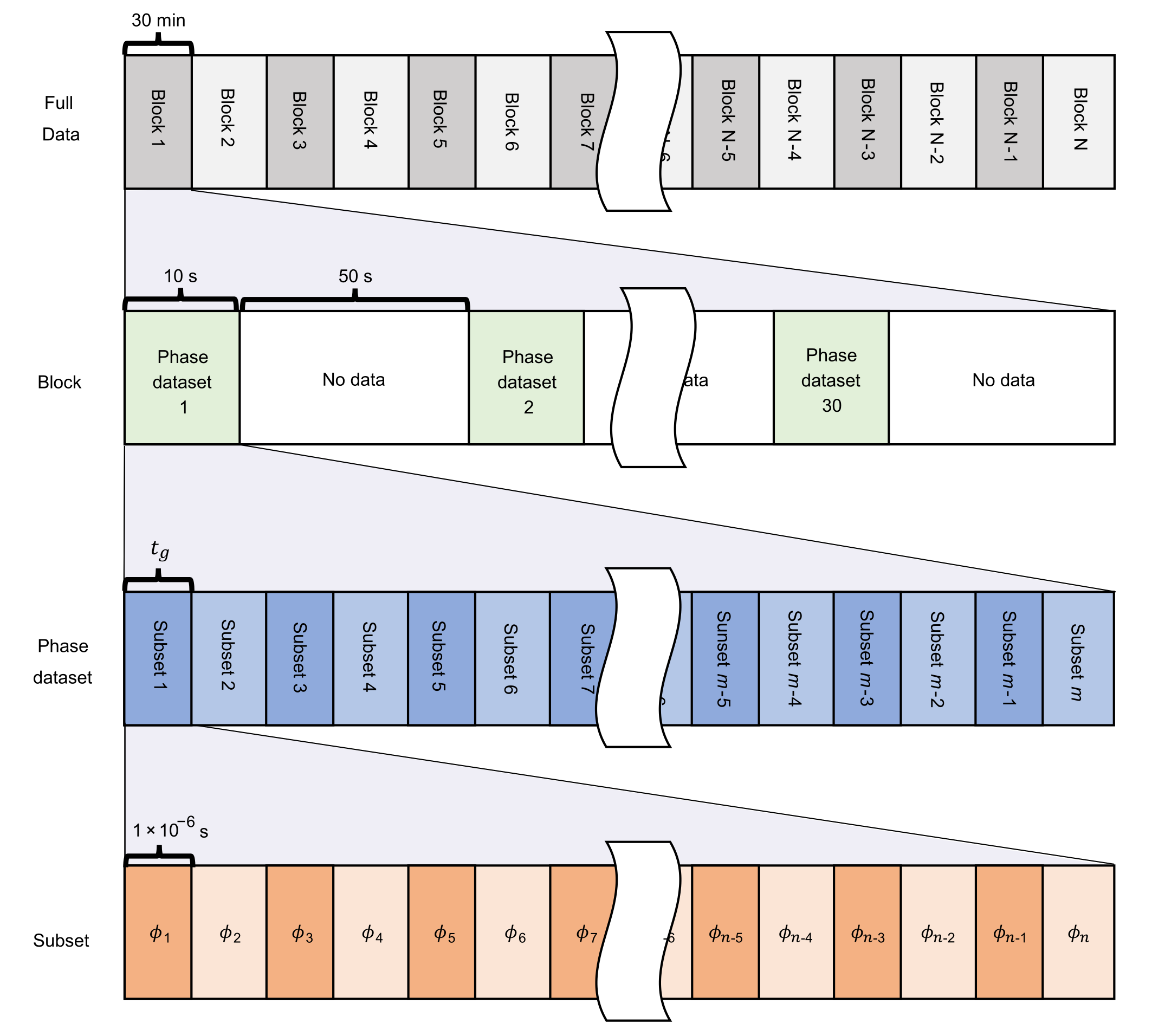}
    \caption{
    {\bf Hierarchy of data structure and procedure of QBER analysis.}
    The entire data is composed of $N$ blocks, each of which corresponds to a duration of $30$~minutes.
    One block is composed of $30$~elements of phase dataset, each of which has a temporal data of $10$~s.
    Before QBER analysis, each phase dataset is split into $m$ subsets so that each subset has a duration of $t_{\rm g}$.
    Then, QBER is computed for each subset from the elementary data $\{\phi_i\}$, where $i = 1, 2,..., n$.
    QBER of one phase dataset is averaged over its subsets.
    For the analysis of long-term fluctuation, QBER is computed for each block by averaging over $30$ phase datasets.
    }
    \label{supl:qber_analysis}
\end{figure}

\end{document}